\documentclass[prd,aps,preprintnumbers, showpacs, nofootinbib,superscriptaddress,notitlepage]{revtex4-1}
\usepackage{amssymb, amsthm}
\usepackage{amsmath}    
\usepackage{mathrsfs}   
\usepackage{graphicx}   
\usepackage{color}      
\usepackage{footnote}   
\usepackage{slashed}

\usepackage{slashed}    
\usepackage{subfigure}  
\usepackage{hyperref}   

\usepackage{rotating}   
\usepackage{multirow}   
\usepackage[normalem]{ulem} 
\usepackage[utf8]{inputenc}

\newcommand{\beq}{\begin{equation}}
\newcommand{\eeq}{\end{equation}}

\begin{document}

\preprint{}

\title{Hadronic molecule interpretation of $T^+_{cc}$ and its beauty-partners  }

\author{Huimin Ren}
\author{Fan Wu}
\author{Ruilin Zhu}
\email{Corresponding author:rlzhu@njnu.edu.cn}
\affiliation{Department of Physics and Institute of Theoretical Physics,
Nanjing Normal University, Nanjing, Jiangsu 210023, China}

\date{\today}

\begin{abstract}
Motivated by the latest discovery of a new tetraquark $T_{cc}^+$ with two charm quarks and two light antiquarks by LHCb Collaboration, we investigated the $DD^*$ hadronic molecule interpretation of $T_{cc}^+$. By calculation, the mass and the decay width of this new structure $T_{cc}^+$ can be understood in one-meson
exchange potential model. The binding energies for these $DD^*$ hadronic molecules with $J^P=1^+$ are around $1MeV$. Besides, we also studied the possible beauty-partners $T_{bb}(10598)$ of  hadronic molecule $T^+_{cc}$, which may be feasible
in future LHCb experiments.
\end{abstract}

\maketitle

\section{Introduction}
\label{SEC:Introduction}
Hadron spectroscopy provides a unique window for us to understand the fundamental strong interactions.
In naive quark model, hadrons are established by quark-antiquark pair or three quarks objects.
However, the Quantum Chromodynamics (QCD) theory tells us that some exotic states
 such as multiquark states or gluon-participated states, which are apart from the conventional configurations,
may also be confined into a color-singlet hadron. The earliest evidence of exotic states is
the X(3872) discovered by the Belle Collaboration in 2003, which lies above the two open charm meson
threshold but has a very narrow decay width ($\Gamma <
1.2MeV$)~\cite{Belle:2003nnu}. Interpretation and verification of the special properties of exotic states attracted
a lot of attempt from both theoretical and experimental aspects (see the reviews~\cite{Chen:2016qju,Guo:2017jvc,Yamaguchi:2019vea}).
The studies of exotic states are not only to gradually filling in the period table of hadrons, but also to enriching our knowledge
of QCD color-confining principle.

Very recently, the LHCb Collaboration have reported the first discovery of a new tetraquark $T_{cc}^+$ with two charm quarks and two light antiquarks
in the $D^0D^0\pi^+$ mass spectrum using a proton-proton collision data set corresponding to an integrated luminosity of $9fb^{-1}$~\cite{LHCb:2021vvq},
where the mass and decay width of $T_{cc}^+$ are determined as
\begin{align}
\delta_m=m_{T_{cc}^+}-(m_{D^{*+}}+m_{D^0})=&-273\pm61\pm5_{-14}^{+11} keV,~~~~
\Gamma_{T_{cc}^+}=410\pm165\pm43_{-38}^{+18} keV.\nonumber
\end{align}
And the spin-parity is determined as $J^P=1^+$. Later the LHCb Collaboration have released a more
profound decay analysis~\cite{LHCb:2021auc}, then the mass and decay width of $T_{cc}^+$ are updated as
\begin{align}
\delta_m=m_{T_{cc}^+}-(m_{D^{*+}}+m_{D^0})=&-361\pm40 keV,~~~~
\Gamma_{T_{cc}^+}=47.8\pm1.9 keV.\nonumber
\end{align}

  The LHCb exotic state $T_{cc}^+$ has an electrical charge and two charm quantum numbers and thus leads to a strong evidence of
least quark content $[cc\bar{u}\bar{d}]$. This exotic system has interesting points. First the two heavy quarks inside the system have small relative velocities due
to its large masses compared to the QCD typical energy scale ($m_Q\gg \Lambda_{QCD}$). There exists an attractive color force between
 the color antitriplet heavy  quark pair. Similar attraction is produced for the light antiquark pair. From these arguments, diquark models were employed~\cite{Carlson:1987hh,Navarra:2007yw,Lee:2009rt,Feng:2013kea,Yan:2018gik,He:2020jna,Xing:2019hjg,He:2016xvd,Wang:2016tsi}.
 On the other hand,  a lower bound state may be produced between two heavy hadrons by exchanging light mesons. Hadronic molecules are also
 popular choices for the system of two heavy quarks and two light antiquarks~\cite{Tornqvist:1993ng,Janc:2004qn,Carames:2011zz,Liu:2019stu,Dong:2021bvy,Meng:2021jnw,Chen:2021vhg}.
 In addition, there are other proposals to explain the doubly heavy tetraquarks: Compact tetraquarks~\cite{Gelman:2002wf,Vijande:2007rf,Ebert:2007rn,Ikeda:2013vwa,Luo:2017eub,Karliner:2017qjm,Eichten:2017ffp,Wang:2017uld,Cheung:2017tnt,Richard:2018yrm,Park:2018wjk,Francis:2018jyb,Junnarkar:2018twb,Deng:2018kly,Yang:2019itm,Cheng:2020wxa,Lu:2020rog,Braaten:2020nwp,Gao:2020ogo,Faustov:2021hjs,Xin:2021wcr,Guo:2021yws,Agaev:2021vur,Sindhu:2021wlp,Lu:2021kut,An:2020jix,An:2020vku,Qin:2020zlg},
Chiral quark model~\cite{Pepin:1996id,Vijande:2003ki,Tan:2020ldi}, Constituent Quark Model~\cite{Noh:2021lqs} and Hydrogen-like molecules~\cite{Zhu:2019iwm}. The production properties of  doubly heavy tetraquarks have been studied in literatures, for example Refs.~\cite{Ali:2018xfq,Ali:2018ifm,Qin:2021dqo,Huang:2021urd}, while the decay properties of
doubly heavy tetraquark have been studied  in literatures~\cite{Ling:2021bir,Feijoo:2021ppq,Fleming:2021wmk,Xing:2018bqt,Hernandez:2019eox,Agaev:2018khe,Agaev:2020mqq,Agaev:2019qqn,Agaev:2020dba}.

Considering that the LHCb exotic state $T_{cc}^+$ is near threshold of a pair of charm mesons, we will investigate  the $DD^{\ast}$ hadronic molecule interpretation of $T_{cc}^{+}$ in this work. From the mass of $T_{cc}^{+}$,
it is extremely close to the $DD^{\ast}$ threshold. The binding energy is less than $1MeV$. We will study
the mass and decays of doubly charm tetraquarks in one-meson
exchange potential (OMEP) model. The effective coupling constants among light mesons and charm mesons are revisited. By power counting,
we only consider the leading order contribution from the lightest mesons, i.e. pseudoscalar mesons. Then the number of parameters is further reduced in OMEP model.
By the investigation of the decay channels, it is also possible to hunting for   the charge-parters $T_{cc}^0$ and
$T_{cc}^{++}$  states.
As a by-product, we also study the mass spectra of the possible doubly bottomed  tetraquarks $T_{bb}$ and discuss their golden decay channels.

This paper is organized as follows. We give the low energy effective Lagrangian and the effective potential after the introduction section.
 The OMEP model is employed to extract the effective potential. In Sec. III,
we present the calculation detail of the mass of the possible ground states of doubly heavy tetraquarks below the heavy meson pair $HH^*$ threshold. In Sec. IV, we give the decay amplitude and decay width
of the process $T_{QQ} \to H+H+\pi$. We conclude in the end.

\section{Low momentum interaction effective theory}
In the heavy quark limit $m_Q\to \infty$, the heavy quark behaves like a static point and  the heavy meson dynamics is determined by
the degree of freedom of the light quark. In heavy quark spin symmetry, the spin-0 and spin-1 heavy-light mesons are combined into
a $4\times4$ matrix
\begin{equation}
{\cal H}_{a}=\frac{1+\slashed{v}}{2}\left[H_{a}^{* \mu} \gamma_{\mu}+iH_{a} \gamma_{5}\right],
\end{equation}
where the pseudoscalar and vector heavy-light meson fields are explicitly expressed as $H_a=(D^0,D^+,D_s^0)$ and $H^*_a=(D^{*0},D^{*+},D_s^{*0})$
for charm sector, $H_a=(\bar{B}^-,\bar{B}^0,\bar{B}_s^0)$ and $ H^*_a=(\bar{B}^{*-},\bar{B}^{*0},\bar{B}_s^{*0})$ for bottom sector; $v$ is the velocity of the heavy quark with the constraint $\slashed{v} {\cal H}_{a}={\cal H}_{a}$.
Here $H^{(*)}_a$ is a triplet in SU(3) flavor symmetry when considering the fact that the masses of light quarks $u$, $d$, and $s$
 can be ignored  compared with the heavy quark mass $m_Q$.

When one considers the exchanging of low energy light mesons between heavy hadrons, it is required to employ the Chiral perturbation theory.
Using this low energy theory it becomes easy to separate the long and short range dynamics.
The doubly charm tetraquark $T_{cc}^{+}$ observed in LHCb experiment is very close to the threshold of $D D^*$,
the binding energy is less than $1MeV$ if we treat the $T_{cc}^{+}$ as a $D D^*$ bound state.
We expect that the low energy expansion converges well.
For the decay channel $T_{cc}^{+}\to D^0+D^0+\pi^+$, the final particles $D^0D^0\pi^+$
will have small velocities  and can be treated as nonrelativistic objects. For example, the maximum energy of $D^0$ in
$T_{cc}^+ \to D^0D^0\pi^+$ is $E^{max}_{D^0}=\frac{m_{T_{cc}}^2+m_D^2-(m_D+m_\pi)^2}{2m_{T_{cc}}}=1867.68MeV$, which is only
2.84MeV above the mass of $D^0$ meson with $m_{D^0}=1864.84MeV$. Similarly, the maximum energy of $\pi^+$ in $T_{cc}^+ \to D^0D^0\pi^+$ is $E^{max}_{\pi^+}=\frac{m_{T_{cc}}^2+m_\pi^2-(m_D+m_D)^2}{2m_{T_{cc}}}=144.85MeV$, which is only
5.279MeV above the mass of $\pi^+$ meson with $m_{\pi^+}=139.57MeV$. Thus we also use the low-energy effective theory to study
the $T_{cc}^{+}\to D^0+D^0+\pi^+$  decay properties.

The low momentum interaction effective theory Lagrangian at leading order is written as~\cite{Stewart:1998ke}
\begin{eqnarray}
\mathcal{L}_{0}=\frac{f_P^{2}}{8} \operatorname{Tr} \partial^{\mu} \Sigma \partial_{\mu} \Sigma^{\dagger}
-i\operatorname{Tr} \bar{{\cal H}}_{a}  v_\mu(\delta_{a b} \partial^{\mu}+iV_{a b}^{\mu}) {\cal H}_{b}+g_P \operatorname{Tr} \bar{{\cal H}}_{a} {\cal H}_{b} \gamma_{\mu} \gamma_{5} A_{b a}^{\mu},
\end{eqnarray}
where  $V_{a b}^{\mu}=\frac{1}{2}\left(\xi^{\dagger} \partial^{\mu} \xi+\xi \partial^{\mu} \xi^{\dagger}\right)_{a b}$ and $A_{a b}^{\mu}=\frac{i}{2}\left(\xi^{\dagger} \partial^{\mu} \xi-\xi \partial^{\mu} \xi^{\dagger}\right)_{a b}$ and $\xi=\sqrt{\Sigma}$.
$\Sigma=\mathrm{Exp}(2iM/{f_P})$ is exponentially related to the light
pseudoscalar mesons with
\begin{eqnarray}
 M=\begin{pmatrix}
 \frac{\pi^0}{\sqrt{2}}+\frac{\eta}{\sqrt{6}}  &\pi^+ & K^+\\
 \pi^-&-\frac{\pi^0}{\sqrt{2}}+\frac{\eta}{\sqrt{6}}&{K^0}\\
 K^-&\bar K^0 &-2\frac{\eta}{\sqrt{6}}
 \end{pmatrix}.
\end{eqnarray}

In principle, the vector and scalar light mesons may also bring new effects in the binding and decays properties of the
near-threshold doubly charm tetraquark states in OMEP model, but these effects are expected to be suppressed as ${\cal O }(m^2_\pi/m^2_{V,S})$ according
to the power counting rules. Compared to the long distance interaction from  pseudoscalar light mesons, the interactions from  scalar and vector light mesons are medium and short ranges. On the other hand, one needs to introduce more parameters in the  effective theory and
some of them are not well investigated currently. We will
address these points in future works.

Using the above Lagrangian, the two-body potential between a vector and a pseudoscalar heavy mesons becomes
as~\cite{Tornqvist:1993ng,Yamaguchi:2019vea}
\begin{eqnarray}
V_{H H^*}(\vec{q})=-\frac{g^2_P}{f^2_P} \frac{\vec{\tau_1}\cdot\vec{\tau_2}\vec{\varepsilon_1}\cdot\vec{q}\vec{\varepsilon^*_2}\cdot\vec{q}}{\vec{q}^2+\mu^2} .
\end{eqnarray}

\section{Doubly heavy tetraquark spectra from $H H^*$ system}
In this section, we will investigate the spectra of doubly heavy tetraquark from the possible $H H^*$ bound states in
OMEP model. After implementing the Fourier transformation on the potential in momentum space, the potential in coordinate space can be obtained.
Considering the size of the exchanged light meson, the Fourier transformation on the potential with the dipole form factors becomes
\begin{eqnarray}
V_{H H^*}(r)=\int\frac{d^3 \vec{q}}{(2\pi)^3}e^{i\vec{q}\cdot\vec{r}}V_{H H^*}(\vec{q})\left(\frac{\Lambda^2-\mu^2}{\Lambda^2+\vec{q}^2}\right)^2,
\end{eqnarray}
where a UV cut-off $\Lambda$ is introduced to regularize the short-distance effects.

In coordinate space, the one-meson exchange potential is then written as~\cite{Tornqvist:1993ng,Yamaguchi:2019vea}
\begin{eqnarray}
V_{H H^*}(r)=-(\frac{g_P}{f_P})^2\gamma_I\left(C_0 C(r)+S_{12}(\hat{r})T(r)\right),
\end{eqnarray}
where
\begin{align}
S_{12}(\hat{r})=&3(\vec{\varepsilon}_1\cdot \hat{r})(\vec{\varepsilon^*}_2\cdot \hat{r})-1,\\
C(r)=&\frac{\mu^{2}}{4 \pi}\left(\frac{\mathrm{e}^{-\mu r}}{r}-\frac{\mathrm{e}^{-\Lambda r}}{r}-\frac{\left(\Lambda^{2}-\mu^{2}\right)}{2 \Lambda} \mathrm{e}^{-\Lambda r}\right),\\
T(r)=& \frac{1}{4 \pi}\left(\left(3+3 \mu r+\mu^{2} r^{2}\right) \frac{\mathrm{e}^{-\mu r}}{r^{3}}-\left(3+3 \Lambda r+\Lambda^{2} r^{2}\right) \frac{\mathrm{e}^{-\Lambda r}}{r^{3}}+\frac{\mu^{2}-\Lambda^{2}}{2}(1+\Lambda r) \frac{\mathrm{e}^{-\Lambda r}}{r}\right).
\end{align}
In the potential the scale is chose as
$\mu^2=m_P^2-(m_{D^*}-m_D)^2$ due to the recoil effect from unequal heavy mesons. For the $D D^*$ interactions, $\mu^2<0$ and we
 should replace $\mu^2$ as $|\mu^2|$ and take the real parts in the integrals~\cite{Yamaguchi:2019vea}.  The form factor parameter $\Lambda$ is chose
as usual $\Lambda\simeq 1.1GeV$. A higher UV cut-off $\Lambda$ may be employed if one includes the medium and short dynamics.
When interpreting the doubly charm tetraquark $T^+_{cc}$ as the possible $D D^*$ bound states with $J^P=1^+$,
the $T^+_{cc}$  state can be rewritten as
\begin{align}
|T^+_{cc}\rangle=&\frac{|D^0 D^{*+}\rangle \pm |D^+ D^{*0}\rangle}{\sqrt{2}}.
\end{align}
 For the $D D^*$  system with $J^P=1^+$, the $D D^*$ may be in the S-wave state with orbital angular momentum $\ell=0$ or D-wave
state  with orbital angular momentum $\ell=2$. The parameter $\gamma_I$ is expressed as $\gamma_I=-2(I(I+1)-I_1(I_1+1)-I_2(I_2+1))/3$ with isospin quantum number $I_i$
of the hadrons. Consider the above mixing between S-wave and D-wave states, one has the matrix~\cite{Tornqvist:1993ng}
\begin{align}<C_0>=\left(\begin{array}{cc}
 1 & 0 \\
 0 & 1
 \end{array}\right),~~<S_{12}(\hat{r})>=\left(\begin{array}{cc}
 0& -\sqrt{2} \\
- \sqrt{2} & 1
 \end{array}\right).
 \end{align}

In nonrelativistic approximation, the $D D^*$ binding energy $E$ can be solved by Schr\"{o}dinger equation
\begin{equation}
\left(-\frac{\hbar^{2}}{2 \mu_{R}} \nabla^{2}+V_{DD^*}(r)\right) \psi(\mathbf{r})=E \psi(\mathbf{r}),
\end{equation}
where $\mu_R=m_{D}m_{D^*}/(m_{D}+m_{D^*})$ is the reduced mass\cite{Zhu:2016arf}. We only focus on the stable bound states with
binding energy $E<0$.

In the calculation, one needs to input the parameter values.  The hadron masses are adopted from PDG~\cite{ParticleDataGroup:2020ssz}: $m_{D^{0}}=1864.84 MeV$, $m_{D^{\pm}}=1869.65 MeV$, $m_{D^{\ast0}}=2006.85MeV$,
$m_{D^{\ast\pm}}=2010.26MeV$,$m_{D^\pm_{s}}=1968.34MeV$,$m_{D_{s}^{*\pm}}=2112.2 MeV$,
	$m_{B^{0}}=5279.65 MeV$,$m_{B^{\pm}}=5279.34 MeV$, $m_{B^0_{s}}=5366.88 MeV$,
	$m_{B^{\ast 0}}=5324.65 MeV$, $m_{B^{\ast\pm}}=5324.65 MeV$, $m_{B^{\ast 0}_{s}}=5415.4 MeV$,
	$m_{\pi^{\pm}}=139.57 MeV$, $m_{\pi^{0}}=134.977MeV$,
	$m_{K^{\pm}}=493.677 MeV$, $m_{K^{0}}=497.611 MeV$, $m_{\eta}=547.862 MeV$. The effective coupling constants $g_P$ are needed to extract
from experiments or lattice QCD calculations. In flavor SU(3) symmetry, they are approximated to be equal to $g_\pi$.
For the $D^*\to D+\pi$  decays, the Feynman amplitude can be written as
\begin{align}
iM(D^*\to  D+\pi)=&i\lambda_\pi p^\mu_\pi \varepsilon_{D^*,\mu}.
\end{align}
The data of the  decay widths and branching ratios can be inputted from PDG~\cite{ParticleDataGroup:2020ssz}:
$\Gamma_{D^{*\pm}}=83.4 keV$, $\Gamma_{D^{*0}}<2.1 MeV$, ${\cal B}(D^{*\pm}\to D^0\pi^\pm)=67.7\%$,
${\cal B}(D^{*\pm}\to D^\pm\pi^0)=30.7\%$, ${\cal B}(D^{*0}\to D^0\pi^0)=64.7\%$. Use the $D^{*\pm}\to D^0\pi^\pm$ channel,
the effective coupling constant is estimated as $\lambda_\pi \sim 16.8$. Use the $D^{*\pm}\to D^+\pi^0$ channel,
the effective coupling constant is estimated as $\lambda_\pi \sim 11.9$. While use the ${\cal B}(D^{*0}\to D^0\pi^0)=64.7\%$
channel and  $\Gamma_{D^{*0}}\approx 60 keV$, one can get $\lambda_\pi \sim 12.3$.  And the effective coupling constant $g_\pi$ is estimated from $\lambda_\pi $  as
\begin{align}
g_\pi \simeq\frac{\lambda_\pi f_\pi}{2\sqrt{m_{D^*}}\sqrt{m_{D}}}\sim [0.4,0.6].
\end{align}
	
We list the  numerical results for the $D D^*$ bound states in Tab.~\ref{MassTQQ}, where the isospin, strange, and beauty partners of
 hadronic molecule $T_{cc}^+$ are also given. The binding energies for $D^0 D^{*+}$ and $D^{*0}D^+$ with $J^P=1^+$ are near to $0.6MeV$.
 The binding energies for strange partners with $J^P=1^+$ are close  to $1MeV$.
While the binding energies for these $BB^*$ hadronic molecules without strange quantum numbers are  around $6MeV$ and then $T_{bb}$ states are more stable.

\begin{table}\caption{Predictions of the masses ({\rm MeV}) of $H H^*$ stable hadronic molecules with
spin-parity $J^P=1^+$. The uncertainty is from the choice of the effective coupling $g_\pi=0.5\mp0.1$. }
{\small \begin{tabular}{|c|c|c|c|c|c|c|c|}\hline\hline
$T_{cc}$ states & Isospin & Contents & Mass({\rm MeV}) & $T_{bb}$ states   & Isospin & Contents & Mass({\rm MeV})\\\hline
$T^+_{cc}$ &0&$D^0 D^{*+},~D^{*0}D^+$ &$3875.1^{+0.2}_{-0.3}$ &$T^-_{bb}$ &0&$\bar{B}^0 \bar{B}^{*-},~\bar{B}^{*0}\bar{B}^-$ &$10598^{+2}_{-3}$
\\\hline
$T^0_{cc}$ &0&$D^0 D^{*0}$ &$3871.0^{+0.2}_{-0.3}$ &$T^0_{bb}$ &0&$\bar{B}^0 \bar{B}^{*0}$ &$10598^{+2}_{-3}$
\\\hline
$T^{++}_{cc}$& 0&$D^+ D^{*+}$ &$3879.2^{+0.2}_{-0.3}$ &$T^{--}_{bb}$& 0&$\bar{B}^- \bar{B}^{*-}$ &$10598^{+2}_{-3}$
\\\hline
$T^{s,+}_{cc}$ &$\frac{1}{2}$&$D^0 D_s^{*+},~D^{*0}D_s^+$ &$3974.8^{+0.4}_{-0.5}$ &$T^{s,-}_{bb}$& $\frac{1}{2}$&$\bar{B}_s^0 \bar{B}^{*-},~\bar{B}_s^{*0}\bar{B}^-$ &$10692.4^{+0.3}_{-0.4}$
\\\hline
$T^{s,++}_{cc}$ &$\frac{1}{2}$&$D_s^+ D^{*+},~D_s^{*+ }D^{+}$ &$3979.3^{+0.2}_{-0.3}$ &$T^{s,0}_{bb}$ &$\frac{1}{2}$&$\bar{B}_s^0 \bar{B}^{*0},~\bar{B}_s^{*0}\bar{B}^0$ &$10692.4^{+0.3}_{-0.3}$
\\\hline
$T^{ss,++}_{cc}$ &0&$D_s^+ D_s^{*+}$ &$4079.6^{+0.2}_{-0.2}$ &$T^{ss,0}_{bb}$& 0&$\bar{B}_s^0\bar{B}_s^{*0}$ &$10781.7^{+0.3}_{-0.2}$
\\\hline
\hline\hline
\end{tabular}}\label{MassTQQ}
\end{table}

\section{ $T_{QQ}\to H H \pi$  decays}

\begin{figure}[th]
\begin{center}
\includegraphics[width=0.5\textwidth]{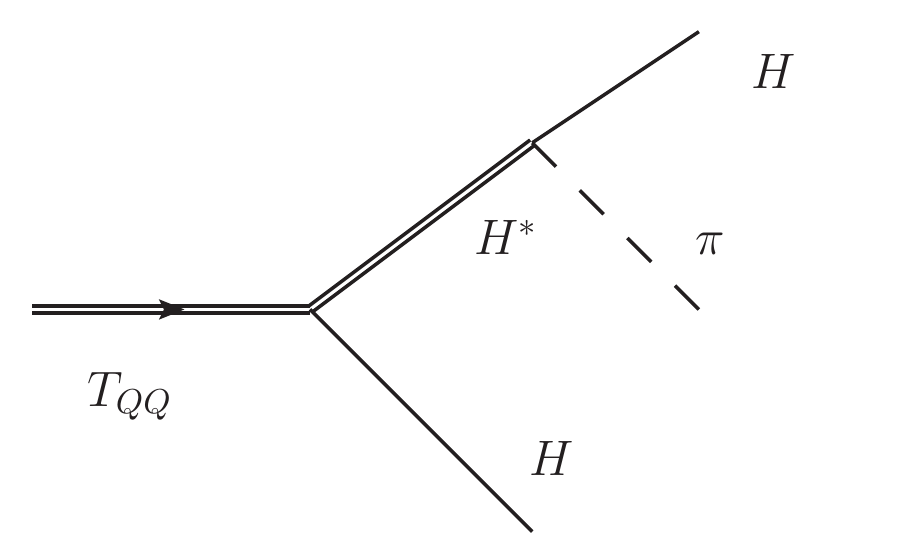}
\caption{ Typical Feynman diagram  for  the $T_{QQ}\to H H \pi$  decay.}\label{fig:Tcc}
\end{center}
\end{figure}

In this section, we will study the decays of $T_{QQ}\to H H \pi$. Here we only focus on the $H H^*$ stable hadronic molecules with
spin-parity $J^P=1^+$ given  in Tab.~\ref{MassTQQ}. Consider the fact that the mass splitting between $D^*$ and $D$ mesons is larger than
 the pion mass, there have $D^*\to D+\pi$ decay channels.  $T_{cc}\to D D\pi$ is also allowed when $T_{cc}$ is close to or above $D D^*$ threshold. However, the mass splitting between $B^*$ and $B$ mesons is small and less than
 the pion mass. Thus $T_{bb}\to B B \pi$ channel is  forbidden due to the lack of  phase space when $T_{bb}$ is below or close to $BB^*$ threshold.  The LHCb collaboration have employed the golden channel $T^+_{cc}\to D^0 D^0\pi^+$ in the discovery of the first doubly charm
tetraquark. The typical Feynman diagram is  plotted in Fig.~\ref{fig:Tcc}.

For $T^+_{cc}\to D^0 D^0\pi^+$ and $T^+_{cc}\to D^0 D^+\pi^0$ processes, the leading-order Feynman amplitudes are similar and can be written as
\begin{align}
iM(T^+_{cc}\to D D \pi)=&i\lambda_\pi p^\mu_\pi\frac{-i}{(p_{D}+p_\pi)^2-m_{D^*}^2}(-g_{\mu\nu}+\frac{(p_{D}+p_\pi)_{\mu}(p_{D}+p_\pi)_{\nu}
}{(p_{D}+p_\pi)^2})im_{T_{cc}}\lambda_{T_{cc}} \varepsilon_{T_{cc}}^\nu,
\end{align}
where  $\lambda_{T_{cc}}$ is the effective coupling constant for the $T_{cc}-D-D^*$ vertex.

In general, the three-body partial decay width is written as~\cite{Chen:2012ju}
\begin{eqnarray}
\frac{d\Gamma}{d s dt}=\frac{1}{(2\pi)^3}\frac{1}{32m_{T_{cc}}^3}|\bar{M}|^2\; ,\end{eqnarray}
where $s$ represents the invariant mass of two heavy charm mesons and $t$ represents the invariant mass of one heavy charm meson
and the pion. For the process $T^+_{cc}\to D^0 D^+\pi^0$, the
phase space constraints then read as~\cite{Qiao:2011zc}
\begin{eqnarray} t_{min}=(E_2+E_3)^2-(\sqrt{E_2^2-m_{D^+}^2}+
\sqrt{E_3^2-m_{\pi^0}^2})^2\; ,t_{max}=(E_2+E_3)^2-(\sqrt{E_2^2-m_{D^+}^2}-
\sqrt{E_3^2-m_{\pi^0}^2})^2\; ,\
 \end{eqnarray}
and
\begin{eqnarray} s_{min}=(m_{D^+}+m_{D^0})^2\;, s_{max} = (m_{T_{cc}}-m_{\pi^0})^2\; ,\  . \end{eqnarray}
The energies in the $s$ rest frame are
\begin{eqnarray}E_2 = \frac{s-m_{D^0}^2+m_{D^+}^2}{2\sqrt{s}}\; ,\ E_3 = \frac{m_{T_{cc}}^2-s-m_{\pi^0}^2}{2\sqrt{s}}\; . \end{eqnarray}

The decay widths  of $T_{QQ}\to H H \pi$ can be estimated as
\begin{align}
\Gamma(T^+_{cc}\to D^0 D^+\pi^0)\approx \left(\frac{\lambda_{T_{cc}}}{6}\right)^2\left(\frac{g_\pi}{0.4}\right)^2 \left(50keV\right),\\
\Gamma(T^+_{cc}\to D^0 D^0\pi^+)\approx \left(\frac{\lambda_{T_{cc}}}{6}\right)^2\left(\frac{g_\pi}{0.4}\right)^2 \left(287keV\right),\\
\Gamma(T^0_{cc}\to D^0 D^0\pi^0)\approx \left(\frac{\lambda_{T_{cc}}}{6}\right)^2\left(\frac{g_\pi}{0.4}\right)^2 \left(223keV\right),\\
\Gamma(T^{++}_{cc}\to D^0 D^+\pi^+)\approx \left(\frac{\lambda_{T_{cc}}}{6}\right)^2\left(\frac{g_\pi}{0.4}\right)^2 \left(103keV\right).
\end{align}
Currently it is not easy to determine the value of the effective coupling $\lambda_{T_{cc}}$. If we choose $\lambda_{T_{cc}}=6$ and $g_\pi=0.4$,
the decay width of $T^+_{cc}$ is estimated as $\Gamma(T^+_{cc})\sim 337keV$, which is consistent to the first round data at LHCb experiment.
While we employ the second round LHCb data, the effective coupling $\lambda_{T_{cc}}$ is extracted as $\lambda_{T_{cc}}\simeq 2.26$. In this case,
the decay width of $T^+_{cc}$ is then estimated as $\Gamma(T^+_{cc})\sim [47.8,107.6]keV$ with the choice of the effective coupling $g_\pi=[0.4,0.6]$.
Due to limited phase space, $T^{++}_{cc}$ does not decay into $D^+ D^+\pi^0$. For the strange parters of $T_{cc}^+$, we will discuss its decay properties in future works due to the lack of information of the effective couplings.   For the beauty parters $T_{bb}$, one can use the
$T_{bb}\to B B \gamma$ electromagnetic decay channels to detect due to the limited phase space. The beauty parters $T_{bb}$ shall be more stable than the $T_{cc}$ states in turn.

\section{Conclusion}
\label{SEC:conclusion}
In this paper, we investigated the mass spectrum and the decay properties of the $T_{cc}^+$ state first observed at LHCb experiment.
Numerical results indicate that the $T_{cc}^+$ state can be well-understood in the $D D^*$ hadronic molecule model.
As a byproduct, the mass spectra of the doubly charm tetraquarks and doubly bottomed tetraquarks in heavy $H H^*$ hadronic molecule
framework are studied, some of which may be hunted in future LHCb experiments. Especially, it is worthwhile to notice the stable $J^P=1^+$ doubly bottomed tetraquarks
 $T^-_{bb}(10598)$, $T^0_{bb}(10598)$, and $T^{--}_{bb}(10598)$ from the  $T_{bb}\to B B\gamma$ decay channels.

\acknowledgments

This work is supported by NSFC under grants No.~12075124 and 11975127, and the Youth Talent Support Program of Nanjing Normal University.

\end{document}